\documentclass{wscpaperproc}
\usepackage{latexsym}
\usepackage{graphicx}
\usepackage{mathptmx}
\usepackage[T1]{fontenc}
\usepackage[ruled, noend]{algorithm2e}
\usepackage{booktabs}
\usepackage{multirow}
\usepackage{cite}

\usepackage{amsmath}
\usepackage{amsfonts}
\usepackage{amssymb}
\usepackage{amsbsy}
\usepackage{amsthm}
\usepackage{xcolor}

\usepackage[pdftex,colorlinks=true,urlcolor=blue,citecolor=black,anchorcolor=black,linkcolor=black]{hyperref}

\newtheoremstyle{wsc}
{3pt}
{3pt}
{}
{}
{\bf}
{}
{.5em}
{}

\theoremstyle{wsc}

\newtheorem{proposition}[]{Proposition}
\begin{document}

%
%

\pagestyle{fancyplain}

\thispagestyle{plain}
\firstPageHead{}

\chead{\fancyplain{}{\itshape Zhao, Luo, Xie, and Bai}}

\rhead{}
\cfoot{}
\renewcommand{\headrulewidth}{0pt} 

\makeatletter
\let\@internalcite\cite
\def\cite{\def\@citeseppen{-1000}%
    \def\@cite##1##2{(##1\if@tempswa , ##2\fi)}%
    \def\citeauthoryear##1##2##3{##1 ##3}\@internalcite}
\def\citeNP{\def\@citeseppen{-1000}%
    \def\@cite##1##2{##1\if@tempswa , ##2\fi}%
    \def\citeauthoryear##1##2##3{##1 ##3}\@internalcite}
\def\citeN{\def\@citeseppen{-1000}%
    \def\@cite##1##2{##1\if@tempswa, ##2)\else{}\fi}%
    \def\citeauthoryear##1##2##3{##1 (##3)}\@citedata}
\def\citeA{\def\@citeseppen{-1000}%
    \def\@cite##1##2{(##1\if@tempswa , ##2\fi)}%
    \def\citeauthoryear##1##2##3{##1}\@internalcite}
\def\citeANP{\def\@citeseppen{-1000}%
    \def\@cite##1##2{##1\if@tempswa , ##2\fi}%
    \def\citeauthoryear##1##2##3{##1}\@internalcite}
\def\shortcite{\def\@citeseppen{-1000}%
    \def\@cite##1##2{(##1\if@tempswa , ##2\fi)}%
    \def\citeauthoryear##1##2##3{##2 ##3}\@internalcite}
\def\shortciteNP{\def\@citeseppen{-1000}%
    \def\@cite##1##2{##1\if@tempswa , ##2\fi}%
    \def\citeauthoryear##1##2##3{##2 ##3}\@internalcite}
\def\shortciteN{\def\@citeseppen{-1000}%
    \def\@cite##1##2{##1\if@tempswa, ##2\else{}\fi}%
    \def\citeauthoryear##1##2##3{##2 (##3)}\@citedata}
\def\shortciteA{\def\@citeseppen{-1000}%
    \def\@cite##1##2{(##1\if@tempswa , ##2\fi)}%
    \def\citeauthoryear##1##2##3{##2}\@internalcite}
\def\shortciteANP{\def\@citeseppen{-1000}%
    \def\@cite##1##2{##1\if@tempswa , ##2\fi}%
    \def\citeauthoryear##1##2##3{##2}\@internalcite}
\def\citeyear{\def\@citeseppen{-1000}%
    \def\@cite##1##2{(##1\if@tempswa , ##2\fi)}%
    \def\citeauthoryear##1##2##3{##3}\@citedata}
\def\citeyearNP{\def\@citeseppen{-1000}%
    \def\@cite##1##2{##1\if@tempswa , ##2\fi}%
    \def\citeauthoryear##1##2##3{##3}\@citedata}
%
%
%
\def\@citedata{%
    \@ifnextchar [{\@tempswatrue\@citedatax}%
                  {\@tempswafalse\@citedatax[]}%
}

\def\@citedatax[#1]#2{%
\if@filesw\immediate\write\@auxout{\string\citation{#2}}\fi%
  \def\@citea{}\@cite{\@for\@citeb:=#2\do%
    {\@citea\def\@citea{, }\@ifundefined
       {b@\@citeb}{{\bf ?}%
       \@warning{Citation `\@citeb' on page \thepage \space undefined}}%
{\csname b@\@citeb\endcsname}}}{#1}}%

%
\def\@citex[#1]#2{%
\if@filesw\immediate\write\@auxout{\string\citation{#2}}\fi%
  \def\@citea{}\@cite{\@for\@citeb:=#2\do%
    {\@citea\def\@citea{; }\@ifundefined
       {b@\@citeb}{{\bf ?}%
       \@warning{Citation `\@citeb' on page \thepage \space undefined}}%
{\csname b@\@citeb\endcsname}}}{#1}}%

%
\def\@biblabel#1{}
\makeatother



\newdimen\bibindent
\bibindent=0.0em
\def\thebibliography#1{\section*{\refname}\list
   {}{\settowidth\labelwidth{[#1]}
   \leftmargin\parindent
   \itemindent -\parindent
   \listparindent \itemindent
   \itemsep 0pt
   \parsep 0pt}
   \def\newblock{}
   \sloppy
   \sfcode`\.=1000\relax}


\setlength{\baselineskip}{12.7pt}

\title{Sensitivity Analysis on Interaction Effects of
Policy-Augmented Bayesian Networks}

\author{\begin{center}Junkai Zhao\textsuperscript{1},  Jun Luo\textsuperscript{1}, Wei Xie\textsuperscript{2}, and Zixuan Bai\textsuperscript{1}\\
[11pt]
\textsuperscript{1}Antai College of Economics and Management, Shanghai Jiao Tong University, Shanghai, China\\
\textsuperscript{2} Department of Mechanical and Industrial Engineering, Northeastern University, Boston, MA, USA\end{center}
}

\maketitle

\vspace{-12pt}

\section*{ABSTRACT}
Biomanufacturing plays an important role in supporting public health and the growth of the bioeconomy. Modeling and studying the interaction effects among various input variables is very critical for obtaining a scientific understanding and process specification in biomanufacturing. In this paper, we use the Shapley-Owen indices to measure the interaction effects for the policy-augmented Bayesian network (PABN) model, which characterizes the
risk- and science-based understanding of production bioprocess mechanisms.
In order to facilitate efficient interaction effect quantification, we propose a sampling-based simulation estimation framework. In addition, to further improve the computational efficiency, we develop a non-nested simulation algorithm with sequential sampling, which can dynamically allocate the simulation budget to the interactions with high uncertainty and therefore estimate the interaction effects more accurately under a total fixed budget setting.

\section{INTRODUCTION}
\label{sec:intro}


The biomanufacturing industry has made substantial progress in producing biotherapeutics to treat serious illnesses such as cancer, autoimmune disorders, and infectious diseases. Despite these advancements, modern biomanufacturing has encountered several critical challenges. The challenges primarily arise from the complex nature of biomanufacturing processes, which are characterized by high levels of complexity, significant variability, and a lack of comprehensive process data. Biotherapeutics are usually produced in living organisms, particularly cells, where the underlying biological processes are highly complicated. The manufacturing process involves a sequence of intricately connected unit operations, which adds complexity to the production procedures. Furthermore, the collection of process data is significantly limited in biomanufacturing settings due to the lengthy experiment time and the customized nature of new biotherapeutics, which hinders the development of a sophisticated biomanufacturing process.


Due to the high variability, measurement error and lack of enough data, sensitivity analysis is employed to guide the process specification and risk control for biomanufacturing \shortcite{xie2023stochastic}. Sensitivity analysis studies the 
fluctuation in a model output caused by changing the model inputs, which can be classified into two classes: local sensitivity analysis and global sensitivity analysis. Local sensitivity analysis focuses on the sensitivity of the output to perturbing
the input around a particular value. However, when the input
is random, it is difficult to determine local sensitivity without knowing which value the input can actually take. On the other hand, global sensitivity analysis quantifies the variability in the output resulting from the uncertainty in the inputs across the whole range of possible input values. It can help identify the most sensitive random factors in the context of biomanufacturing. \shortciteN{sobol1993sensitivity} first introduced the Sobol' indices
to quantify how an input influences an output’s variance. Under the assumption of independence between the input variables, the Sobol' indices compare the conditional variance of the output knowing some of the input variables to the total variance of the output. 
\shortciteN{owen2014sobol} proposed new sensitivity indices called Shapley effects and showed that these new indices have many advantages over the Sobol' indices for dependent inputs. They are now widely used in global sensitivity analysis \shortcite{rabitti2020mortality}. 
To improve the computational efficiency of Shapley effects, \shortciteN{song2016shapley} provided an efficient estimate algorithm with Monte Carlo simulation. 

The aforementioned methods focus on measuring the importance of individual model inputs, without considering the interaction effects of different input variables. However, many interactions exist in the biomanufacturing process, including gene-gene interactions \shortcite{cordell2009detecting}, cell-environment interactions  \shortcite{aijaz2018biomanufacturing} and so on. Some interactions may come from common underlying random factors and the systematic observation error of the sensors.
Understanding the interactions can benefit the scientific understanding of the biomanufacturing process. For example, it allows one to discriminate whether an interaction is synergistic (i.e., two model inputs are related
by positive cooperation) or antagonistic (i.e., two model inputs are related by negative cooperation) \shortcite{diouf2021computation}. The quantity of the interaction effects can also be used to classify parameters and determine which type of studies (i.e., DoE, univariate, or none) should be
performed during process characterization
\shortcite{lkacki2018downstream}. To measure the interaction effects of input variables, \shortciteN{rabitti2019shapley} proposed the Shapley-Owen index, which generalizes the Shapley effects to the context of interaction quantification. They showed that the Shapley-Owen index can provide information about synergistic and antagonistic effects. However, they only test the performance of the Shapley-Owen index for the stochastic model with an analytical variance function and up to three inputs. As the number of inputs increases, the computational cost would become unaffordable.



In this
paper, we consider how to conduct global sensitivity analysis for the interactions among different inputs in biomanufacturing based on the Shapley-Owen index. Our analysis is based on the policy augmented Bayesian network (PABN) proposed by \shortciteN{zheng2023policy}, which can leverage the existing
mechanisms and learn from real-world process data \shortcite{zhao2023policy}.  Building 
on PABN, we introduce the Shapley-Owen based sensitivity analysis to assess the interaction effect. Firstly, as the calculation of the Shapley-Owen index requires massive computational resources, we extend the method in \shortciteN{song2016shapley}
and propose a nested Monte Carlo simulation algorithm for Shapley interaction effect estimation in PABN. Compared to the brute force implementation of the Shapley-Owen index formula, our algorithm can benefit from approximation by sampling and reuse of calculation. Secondly, the sample size for each level of simulation can be hard to be determined without expert knowledge and the error analysis cannot be conducted analytically. To overcome these issues, we propose a non-nested estimator to estimate the Shapley interaction effect, which can facilitate error estimation and can be implemented without the determination of the multilevel sample sizes. In addition, the total number of interactions that need to be quantified grows in the square order of the number of factors, which can be extremely large as the number of factors increases. However, within those interactions, there may be some interactions 
that exhibit a low level of uncertainty or variability. In order to enhance the accuracy of estimation within a specified computational budget, we propose a sequential budget allocation algorithm that adaptively identifies which interaction has the most uncertainty and then allocates that interaction for one additional simulation replication. By allocating the computation budget sequentially, our algorithm can achieve higher accuracy using the same budget as the previous Shapley value estimation method. We summarize our contributions as follows:
\begin{itemize}
    \item We propose a sampling-based global sensitivity analysis framework to quantify the impact of the interactions with the Shapley-Owen index in PABN model. 
    \item We develop a non-nested simulation algorithm, which adopts the non-nested estimators and sequential budget allocation strategy to improve the efficiency.
    \item We demonstrate the performance of our methods in terms of both accuracy and efficiency with empirical study.
\end{itemize}

The paper is organized as follows. In Section~\ref{sec: hybrid model}, we present the general PABN model. We introduce the Shapley interaction effect and the nested estimation procedure for PABN model in Section~\ref{sec: shapeffect}. In Section~\ref{sec: budget allocation}, we develop the sequential budget allocation algorithm. 
Then, we study the empirical performance of two proposed algorithms in Section~\ref{sec: experiment} and conclude this paper in Section~\ref{sec: conclusion}.







\section{General Policy Augmented Knowledge Graph}
\label{sec: hybrid model}


In this section, we introduce the general PABN model in \shortciteN{zheng2023policy} for the biomanufacturing process. A typical biomanufacturing process comprises multiple unit operations. Within these unit operations, two types of interacting factors play pivotal roles: one is critical quality attributes (CQAs), which can be considered as the states of the bioprocess (e.g., the concentrations of bioproduct); the other is 
critical process parameters (CPPs), which can be viewed as the actions of the bioprocess (e.g., the feeding rate). Both of them can significantly influence the system's outputs (e.g., drug quality and productivity). In this paper, we represent the CQAs at period $t$ as $\pmb{s}_t=\left(s^1_t,...,s^{d_s}_t\right)$ and the CPPs at period $t$ as $\pmb{a}_t =\left(a^1_t,...,a^{d_a}_t\right)$, where $H$ is the total planning horizon, $1\le t \le H$. The $n$th component of $\pmb{s}_t$ and $\pmb{a}_t$ are denoted by $\pmb{s}^n_t$ and $\pmb{a}^n_t$, respectively. For simplicity, the dimensions of $\pmb{s}_t$ and $\pmb{a}_t$, namely $d_s$ and $d_a$ respectively, are kept constant throughout this study, although they may be time-dependent in reality.

\begin{table}[!b]
\centering
\caption{List of key symbols.}
\label{tab: notation}
\begin{tabular}{llll}
\toprule
Symbol & Definition & Symbol & Definition \\ \midrule

$\pmb{s}_t$& CQA at period $t$ &
$\pmb{a}_t$& CPP at period $t$  \\
$H$& total horizon &
$d_s$& the dimensions of  $\pmb{s}_t$ \\
$d_a$& the dimensions of  $\pmb{a}_t$ &
$\pmb{e}_{t}$ & uncontrollable factors at period $t$\\
$\pmb{w}$ & model parameters &
$\pmb{\theta}_t$ & policy paramters at period $t$\\
$r_t(\pmb{s}_t, \pmb{a}_t)$ & the reward function at each period $t$&
$\pmb\tau^{(l)}$ & $l$th observed trajectory in experiments\\
$\pi_t$ & the policy at period $t$ &
$Y$ & the cumulative reward\\
$\mathcal{D}$ & historical data &
$\mathcal{I}$ & the set of all inputs\\
$g(\cdot\vert\pmb{w})$ & \multicolumn{1}{l}{value function given $\pmb{w}$}&
$Sh_{i,j}$& \multicolumn{1}{l}{Shapley interaction effect between $i$ and $j$}\\
$\Pi(\mathcal{I})$ & \multicolumn{3}{l}{the set of all permutations of the input set $\mathcal{I}$}\\
$P_{i,j}(\pi)$ & \multicolumn{3}{l}{the set of inputs except for $j$ that precede the element $i$ in the permutation $\pi$}\\
\bottomrule
\end{tabular}
\end{table}

In bioprocesses, kinetic models based on the ordinary or partial differential equations (ODE/PDE) are commonly applied to characterize the dynamics of the system, which means $\pmb s_t$ evolves according to $\frac{ d \pmb{s}_t}{d t} = f_t(\pmb{s}_t,\pmb{a}_t)$. Specifically, the functional form of $f_t(\pmb{s}_t,\pmb{a}_t; \pmb{w}_t)$ is expected to be known, while the parameters of the function $\pmb{w}_t$, which can be estimated from data, are subject to uncertainty. Drawing upon prior knowledge of the existing ODE/PDE-based kinetics, \shortciteN{zheng2023policy} introduced the state transition hybrid model, referred to as,
$\pmb{s}_{t+1}=f_t(\pmb{s}_t,\pmb{a}_t; \pmb{\beta}_t) + \pmb{e}_{t+1}$, 
 where 
 the residual $\pmb{e}_{t+1}$ is a random vector representing the uncontrollable factors. Suppose that the residuals follow a multivariate Gaussian distribution, i.e., 
$\pmb{e}=(\pmb{e}_1^\top, \pmb{e}_2^\top,\ldots,\pmb{e}_H^\top)\sim \mathcal{N}(\pmb{0}, \mathbf{V})$, where $v_{ij}$ in $i$th row and $j$th column of $\mathbf{V}$ denotes covariance between the $i$th component and the $j$th component of $\pmb{e}$. The model parameters can be denoted by $\pmb{w} = \{\{\pmb{\beta}_t\}_{t=1}^{H-1}, \mathbf{V}\}$ 
and the distribution of the entire trajectory $\pmb\tau = (\pmb{s}_1,\pmb{a}_1, \pmb{s}_2, \pmb{a}_2, ..., \pmb{s}_H)$ of the stochastic decision process
(SDP) can be written as $p(\pmb{\tau}) = p(\pmb{s}_1)\prod_{t=1}^{H-1}p(\pmb{s}_{t+1}\mid\pmb{s}_t,\pmb{a}_t)p(\pmb{a}_t\mid \pmb{s}_t)$. 
Given the prior distribution of the model parameters and the historical data $\mathcal{D} = \left\{\pmb\tau^{(l)}\right\}^L_{l=1}$, where $\pmb\tau^{(l)}$ denotes $l$th observed trajectory in experiments. We can derive the posterior distribution of model parameters $p(\pmb{w}\vert\mathcal{D})$, which quantifies the uncertainty in parameter estimation due to the lack of information.

At each period $t$, we employ the policy $\pi_t$ to determine the action. Given the policy parameters $\pmb{\theta}_t$, the state vectors $\pmb{s}_t$ are mapped into the space of all potential action values with $\pmb{a}_t = \pi_t\left(\pmb{s}_t; \pmb{\theta}_t\right)$. The parametric policy $\pi _{\pmb{\theta}}= \left\{\pi_t\right\}^{H-1}_{t=1}$ is the collection of these mappings over the entire planning period and is fully characterized by $\pmb{\theta} = \{\pmb{\theta}_t\}_{t=1}^{H-1}$. Let $r_t(\pmb{s}_t, \pmb{a}_t)$ denote the reward function at each period $t$, the output of interest is the random cumulative reward accrued by adhering to the policy specified by $\pmb{\theta}$ over the planning period, expressed as $r(\pmb{\theta}) = \sum_{t=1}^H r_t(\pmb{s}_t,\pmb{a}_t)$. Some key symbols used in the paper are shown in Table~\ref{tab: notation}.

In this study, the focused input variables of sensitivity analysis are the random factors associated with the states, i.e., the residuals. The output of interest is the cumulative reward $Y=r(\pmb{\theta})$ or future states, i.e., $Y\in\{s^{n}_t, n = 1, \ldots, d_s, t = 1,\ldots,H\}$. 
We study how the variance of the $Y$ is affected by the
interactions of the random inputs with a predetermined policy so as to better understand system performance, quantify risk, or indicate where interaction intervention or
management may be desirable.  Our methods can be extended to analyze how the interactions of the model parameters affect the variance of the output.

\section{Shapley interaction effect estimation}
\label{sec: shapeffect}
In Section~\ref{sec: hybrid model}, we introduced the PABN model that is used to characterize the biomanufacturing process and the focused random factors in the PABN model. In this section, we introduce the framework for how to use the Shapley-Owen index to measure the impact of each interaction among the random factors in PABN on the variance of the final reward.
Due to the high dimension of random factors and the existence of model uncertainty in the PABN model described in Section~\ref{sec: hybrid model}, we propose a sampling-based simulation algorithm for the estimation of the Shapley-Owen index in the PABN model, taking into account both computational tractability and model uncertainty.

\begin{algorithm}[!b]
\linespread{1.0}\selectfont
Input: the number of samples for model parameters $K$, the number of permutations $M$, the outer and inner sample size for the simulation of value function $N_{O}$ and $N_{I}$, the policy parameters $\pmb{\theta}$, the posterior distribution of model parameters $p(\pmb{w}\vert\mathcal{D})$, the set of inputs $\mathcal{\mathcal{I}}$,
$\hat{Sh}_{i,j}  = 0 \text{ for } \{i, j\} \in \mathcal{\mathcal{I}}$.
\\
1. Generate the samples of the marginal contribution and calculate the sum of samples:\\
\For{$k = 1,2,\ldots,K$}{
Sample model parameters $\pmb{w}^{(k)}$ from the posterior distribution $p(\pmb{w}\vert\mathcal{D})$:\\ 
\For{$m=1,2,\ldots,M$}{
Sample the random permutation $\pi^{(m)}$ from $\Pi(\mathcal{I})$;\\
\For{all set $\{i, j\}\subseteq \mathcal{I}$}{
$\hat{\mbox{Sh}}_{i,j}=\hat{\mbox{Sh}}_{i,j}+\hat{g}_{nested}(P_{i,j}(\pi^{(m)})\cup \{i, j\}\vert\pmb{w}^{(k)})  - \hat{g}_{nested}(P_{i,j}(\pi^{(m)})\cup\{i\}\vert\pmb{w}^{(k)})- \hat{g}_{nested}(P_{i,j}(\pi^{(m)})\cup\{j\}\vert\pmb{w}^{(k)})+\hat{g}_{nested}(P_{i,j}(\pi^{(m)})\vert\pmb{w}^{(k)})$.
}}}
2. Calculate the estimation of the Shapley interaction effect:
$\hat{Sh}_{i,j} = \hat{Sh}_{i,j}/(KM) \text{ for } \{i,j\} \in \mathcal{I}$.
\caption{Shapley Interaction Effect Estimation on Random Factors of PABN}\label{agm: svgeneral}
\end{algorithm}	

Let $\mathcal{I}$ denote the set of all inputs, where $\mathcal{I}$ can be the set of random factors $\{e_t^n, n=1,\ldots,d_s, t=1,\ldots,H\}$. 
Suppose that the process model parameters $\pmb{w}$ are given,  to quantify the contribution of each input ${i} \in \mathcal{I}$ to the variance of interested output $Y$, the Shapley value of any input ${i}$ is defined as 
\begin{equation*}
    \mbox{Sh}_i\left(Y\vert \pmb{w}\right)
	=\sum_{\mathcal{U}\subset \mathcal{I}/\{{i}\}}\dfrac{(\vert\mathcal{I}\vert -\vert\mathcal{U}\vert-1)! \vert\mathcal{U}\vert!} {\vert\mathcal{I}\vert!}\big[ g(\mathcal{U}\cup\left\{{i}\right\}\vert\pmb{w}) - g(\mathcal{U}\vert\pmb{w}) \big],
\end{equation*}
where $\vert\mathcal{U}\vert$ is the cardinality of subset $\mathcal{U}\subset \mathcal{I}/\left\{i\right\}$ and $g(\cdot\vert\pmb{w})$ is called the value function associated with a set of inputs, which denotes the variance of the output generated by a subset of inputs $\mathcal{U}$. The detailed form of $g(\cdot)$ will be discussed later. We can see that $g(\mathcal{U}\cup\left\{{i}\right\}\vert\pmb{w}) - g(\mathcal{U}\vert\pmb{w})$ is the marginal contribution of input $i$ to the subsets of inputs $\mathcal{U}$ in terms of the expectation or variance of the output. Therefore, the Shapley value is the weighted average of the marginal contributions of $i$ to all subsets $\mathcal{U}$.
The interaction effect between two factors $i$ and $j$ can be defined following \shortciteN{rabitti2019shapley} as 
\begin{equation}
\begin{aligned}
\mbox{Sh}_{i,j}\left(Y\vert\pmb{w}\right)
&=\sum_{\mathcal{U} \in \mathcal{I}/\{i,j\})}\dfrac{(\vert \mathcal{I}\vert-\vert \mathcal{U}\vert-2)!\vert \mathcal{U}\vert!}{(\vert \mathcal{I}\vert-1)!}\big[ g(\mathcal{U}\cup \{i, j\}\vert\pmb{w})\\
&\quad - g(\mathcal{U}\cup\{i\}\vert\pmb{w}) - g(\mathcal{U}\cup\{j\}\vert\pmb{w})+g(\mathcal{U}\vert\pmb{w}) \big],
\end{aligned}
\label{eq: shape_interaction}
\end{equation}
which is called the Shapley-Owen index and measures the weighted average marginal contribution in the presence of two factors $\{i, j\}$. 


It is important to notice that as $\vert\mathcal{I}\vert$ increases, the number of subset $\mathcal{U}$ increases exponentially. To address this issue, we can rewrite Equation~(\ref{eq: shape_interaction}) as
\begin{equation}
    \mbox{Sh}_{i,j}\left(Y\vert\pmb{w}\right)
	=\sum_{\pi \in \Pi(\mathcal{I})}\dfrac{1}{\vert \mathcal{I}\vert!}\big[ g(P_{i,j}(\pi)\cup \{i, j\}\vert\pmb{w}) - g(P_{i,j}(\pi)\cup\{i\}\vert\pmb{w}) - g(P_{i,j}(\pi)\cup\{j\}\vert\pmb{w})+g(P_{i,j}(\pi)\vert\pmb{w}) \big],
	\nonumber 
\end{equation}
where $\Pi(\mathcal{I})$ is the set of all permutations of the input set $\mathcal{I}$, $P_{i,j}(\pi)$ is the set of inputs except for $j$ that precede the element $i$ in the permutation $\pi$. For instance, if the set of inputs is denoted as $\mathcal{I} = \{1, 2, 3, 4, 5\}$, one permutation of the $\mathcal{I}$ can be $\pi = \{3, 1, 2, 5, 4\}$, then $P_{2,4}(\pi) = \{3, 1\}$. 
We can randomly sample permutations from $\Pi(\mathcal{I})$ and average the incremental marginal contributions $g(P_{i,j}(\pi)\cup \{i, j\}\vert\pmb{w}) - g(P_{i,j}(\pi)\cup\{i\}\vert\pmb{w}) - g(P_{i,j}(\pi)\cup\{j\}\vert\pmb{w})+g(P_{i,j}(\pi)\vert\pmb{w})$ to approximate the real Shapley-Owen index.

\begin{figure}[!t]
\vspace{-1em}
	\centering	\includegraphics[width=0.7\textwidth]{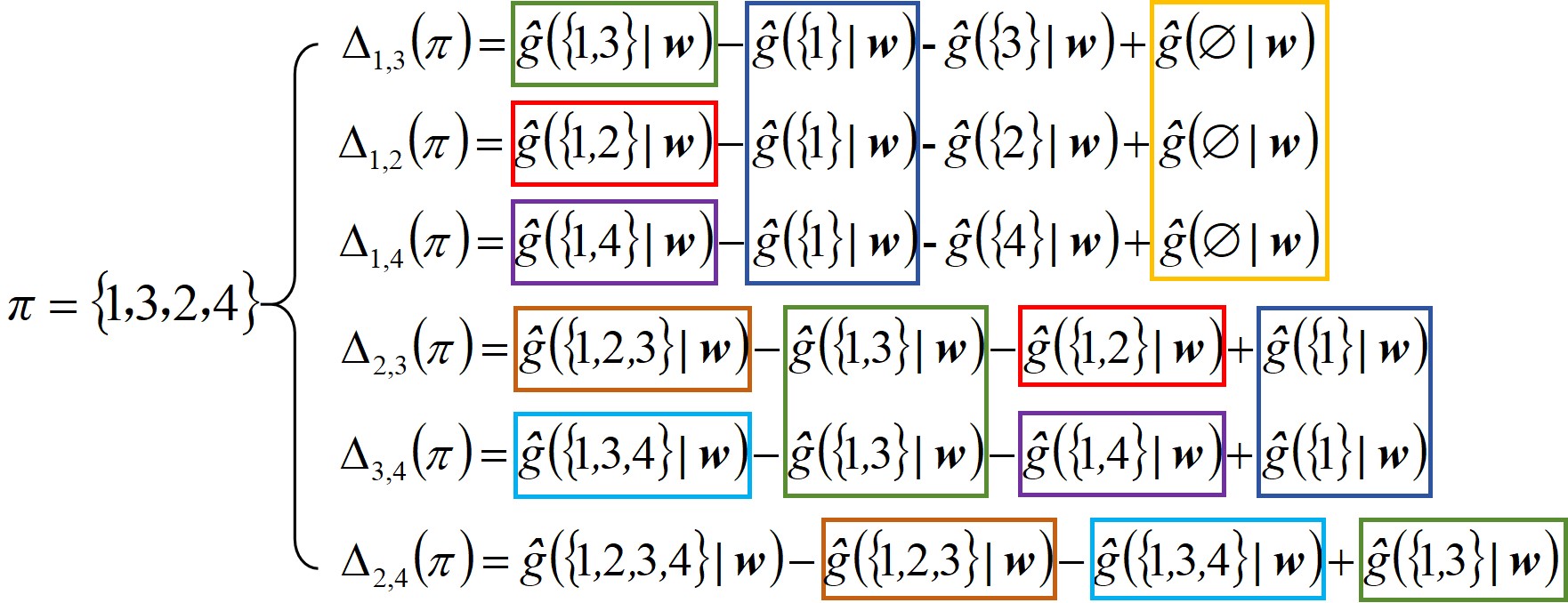}
\caption{An illustration of the reused calculation in Shapley-Owen index estimation for a sampled permutation. Duplicate terms are framed and displayed in the same color.}
	\label{fig: permu}
\end{figure}

Notice that there are some redundant calculations in the Shapley interaction effect estimation. We can record the previous calculation results to reduce computational efforts. For instance, if there are 4 inputs and the sampled permutation $\pi = \{1,3,2,4\}$, the computation process is shown in Figure~\ref{fig: permu}, where $\Delta_{i,j}(\pi) = g(P_{i,j}(\pi)\cup \{i, j\}\vert\pmb{w}) - g(P_{i,j}(\pi)\cup\{i\}\vert\pmb{w}) - g(P_{i,j}(\pi)\cup\{j\}\vert\pmb{w})+g(P_{i,j}(\pi)\vert\pmb{w})$.
We frame the terms that are calculated more than once and the duplicate terms are marked with the same color.
In this example, $\hat{g}(\emptyset|\pmb{w} )$, $\hat{g}(\{1\}|\pmb{w} )$, $\hat{g}(\{1,3\}|\pmb{w} )$, $\hat{g}(\{1,2\}|\pmb{w} )$, $\hat{g}(\{1,4\}|\pmb{w} )$, $\hat{g}(\{1,3,4\}|\pmb{w} )$, and $\hat{g}(\{1,2,3\}|\pmb{w} )$ are calculated more than once. We can record them for subsequent calculations and therefore reduce the computational load by approximately half.

Next, we consider the estimation of the value function. Let $X_{\mathcal{U}}$ be the vector of random inputs included in $\mathcal{U}$. In this study, we use $g(\mathcal{U}) = \mbox{Var}[\mbox{E}[Y\vert X_{\mathcal{U}}]]$
, which can
be interpreted as the component of the output variance that can be explained by a functional
dependence on the input in $\mathcal{U}$. Using the variance decomposition formula, it can be rewritten as $\mbox{Var}[Y] - \mbox{E}[\mbox{Var}[Y\vert X_{\mathcal{U}}]]$, which can be interpreted as the average decrease in output variance when one receives perfect information
about $X_{\mathcal{U}}$. To estimate $g(\mathcal{U})$, \shortciteN{sun2011efficient} introduced the following unbiased nested Monte Carlo estimator: Given the number of outer samples $N_O$ and the number of inner samples $N_I$, let $y_{n_1, n_2}, n_1 = 1, \ldots, N_O, n_2 = 1, \ldots, N_I$ be the overall samples of $Y$, 
which is obtained by firstly sampling $x_{\mathcal{U}}$ from $F_{X_{\mathcal{U}}}$, the joint distribution of random factors in $\mathcal{U}$, then sampling $Y$ from $F_{Y\vert X_{\mathcal{U}} = x_{\mathcal{U}}}$, the conditional distribution of $Y$ given $X_{\mathcal{U}} = x_{\mathcal{U}}$.
Let $\bar{y}_{n_1} = \dfrac{1}{N_I}\sum_{n_2 = 1}^{N_I}y_{n_1, n_2}$, $\bar{\bar{y}} = \dfrac{1}{N_O}\sum_{n_1=1}^{N_O}\bar{y}_{n_1}$, the quantity of interest is estimated by 
\begin{equation*}
    \begin{aligned}
        \hat{g}_{nested}(\mathcal{U}) = \dfrac{1}{N_O - 1}\sum_{n_1=1}^{N_O}(\bar{y}_{n_1} - \bar{\bar{y}})^2 - \dfrac{1}{N_O N_I(N_I-1)}\sum_{n_1=1}^{N_O}\sum_{n_2=1}^{N_I}(y_{n_1,n_2}-\bar{y}_{n_1})^2,
    \end{aligned}
\end{equation*}

The aforementioned
calculation is based on a fixed $\pmb{w}$. To handle the model parameters uncertainty of PABN, we can draw samples $\pmb{w}^{(k)}$, $k = 1\ldots, K$, from the posterior distribution $p(\pmb{w}\vert\mathcal{D})$, then the final estimator of the Shapley-Owen index
is
\begin{equation*}
    \begin{aligned}
        \hat{\mbox{Sh}}_{i,j}&=\dfrac{1}{KM}\sum_{k=1}^{K}\sum_{m=1}^{M}\left[ \hat{g}_{nested}(P_{i,j}(\pi^{(m)})\cup \{i, j\}\vert\pmb{w}^{(k)})  - \hat{g}_{nested}(P_{i,j}(\pi^{(m)})\cup\{i\}\vert\pmb{w}^{(k)}) \right.\\
        &\left.\quad- \hat{g}_{nested}(P_{i,j}(\pi^{(m)})\cup\{j\}\vert\pmb{w}^{(k)})+\hat{g}_{nested}(P_{i,j}(\pi^{(m)})\vert\pmb{w}^{(k)}) \right].
    \end{aligned}
\end{equation*}

The complete statement of the interaction effect estimation procedure is given in Algorithm~\ref{agm: svgeneral}.

\section{Computational Budget Allocation}
\label{sec: budget allocation}
In practice, the calculation of the Shapley interaction effect using Algorithm~\ref{agm: svgeneral} faces several challenges. Firstly,
the implementation of Algorithm~\ref{agm: svgeneral} requires the proper choice of the sample size in each level, i.e., the number of $K$, $M$, $N_O$ and $N_I$, which could be difficult to properly predetermine. Secondly, the complex nested simulation procedure hinders the analysis of the performance guarantee and estimation error of the algorithm. Thirdly, for each sampled permutation, Algorithm~\ref{agm: svgeneral} accumulates the marginal contribution for all subset $\{i,j\}\subseteq \mathcal{I}$, which allocates the total computational effort to each subset $\{i,j\}$ equally. However, if one alternative has very low variance, then it may only require very few samples
to accurately estimate its performance. Therefore, equal budget allocation can be inefficient.
In this section, we provide a non-nested unbiased estimator for the Shapley interaction effect $\mbox{Sh}_{i,j}$, based on which we then propose an adaptive budget allocation procedure for the estimation of $\mbox{Sh}_{i,j}$.

Let $X^{(0)}$ and $X^{(1)}$ be two independent and identically distributed (i.i.d.) random variables generated from $F_X$, $Y^{(1)}$ and $Y^{(2)}$ are i.i.d.\ 
random variables following distribution $F_{Y|X^{(0)}, \pmb{w}}$, the conditional distribution of $Y$ given $X = X^{(0)}$ and model parameter $\pmb{w}$, and $Y^{(3)}$ follows the distribution $F_{Y|X^{(1)}, \pmb{w}}$, the conditional distribution of $Y$ given $X = X^{(1)}$ and model parameter $\pmb{w}$. Then according to \shortciteN{cheng2021nonnested}, $\mbox{Var}[\mbox{E}(Y|X)]=\mbox{E}[Y^{(1)}(Y^{(2)}-Y^{(3)})]$. Therefore, let $x^{(0)}_{\mathcal{U}}$ and $x^{(1)}_{\mathcal{U}}$ be two random samples generated from $F_{X_{\mathcal{U}}\vert\pmb{w}}$, a non-nested unbiased estimator of the value function can be
    $\hat{g}_{nn}(\mathcal{U}\vert \pmb{w}) = y^{(1)}(y^{(2)}-y^{(3)})$,
where $y^{(1)}$ and $y^{(2)}$ are sampled from $F_{Y\vert x^{(0)}_{\mathcal{U}},\pmb{w}}$, $y^{(3)}$ is sampled from $F_{Y\vert x^{(1)}_{\mathcal{U}},\pmb{w}}$. Let $\mbox{Unif}(\Pi(\mathcal{I}))$ denote the uniform distribution over the permutation space $\Pi(\mathcal{I})$. Now we obtain a new estimator of the Shapley interaction effect, which can be written as
\begin{equation*}
\begin{aligned}
    \hat{Sh}_{ij} = \dfrac{1}{N_{i,j}}\sum_{n=1}^{N_{i,j}}\hat{g}_{nn}(P_{i,j}(\pi^{(n)})\cup\{i,j\}\vert \pmb{w}^{(n)})-\hat{g}_{nn}(P_{i,j}(\pi^{(n)})\cup\{i\}\vert \pmb{w}^{(n)})\\-\hat{g}_{nn}(P_{i,j}(\pi^{(n)})\cup\{j\}\vert \pmb{w}^{(n)})+\hat{g}_{nn}(P_{i,j}(\pi^{(n)})\vert \pmb{w}^{(n)})
\end{aligned}
\label{eq: non_nested sv estimator}
\end{equation*}
where $\pi^{(n)}$ and $\pmb{w}^{(n)}$ are sampled from $\mbox{Unif}(\Pi(\mathcal{I}))$ and $p(\pmb{w}\vert\mathcal{D})$, respectively, and $N_{i,j}$ is the sample size used to estimate $Sh_{ij}$.

In Algorithm~\ref{agm: svgeneral}, the computational resources are allocated to each subset $\{i,j\}\subseteq \mathcal{I}$ equally. However, subset $\{i,j\}$ with greater uncertainty in $\hat{Sh}_{i,j}$ may require more samples to reduce the estimation error. As a result, the equal allocation
approach could not efficiently use the simulation resource. To avoid such issue, we develop a sequential procedure that can efficiently employ the simulation budget to estimate $\hat{Sh}_{i,j}$ for each $\{i,j\}\subseteq \mathcal{I}$. As the focus of this paper is the estimation of the true Shapley-Owen indices rather than the identification of the most influential interactions, we use the confidence interval to quantify the estimation uncertainty and to guide the allocation of total simulation budget.

Based on the central limit theorems, we first construct the asymptotic confidence interval for $\hat{Sh}_{i,j}$ as in \shortciteN{cheng2021nonnested}:
\begin{proposition}
\label{thm: CI for sv}
Given the sample size $N_{i,j}$ and the significance level of $\alpha$,
\begin{align*}
\mbox{Pr}\Big\{\hat{Sh}_{i,j} \in \big(\hat{Sh}_{i,j}+z_{\alpha/2}\hat{\sigma}_{i,j}/\sqrt{N_{i,j}}, \ \hat{Sh}_{i,j}-z_{\alpha/2}\hat{\sigma}_{i,j}/\sqrt{N_{i,j}} \big)\Big\} \simeq 1-\alpha,
\end{align*}
where $z_{\alpha}$ is the $\alpha/2$ quantile of the standard normal distribution, $\hat{\sigma}_{i,j}$ is the sample standard deviation, and $\simeq$ indicates the asymptotic equality as $N_{i,j}  \to \infty$. 
\end{proposition}

\begin{algorithm}[!t]
\linespread{1.0}\selectfont
Input: the number of total iterations $N$, the number of iterations for Stage~1 $N_0$,
the policy parameters $\pmb{\theta}$, the posterior distribution of model parameters $p(\pmb{w}\vert\mathcal{D})$, the set of inputs $\mathcal{\mathcal{I}}$,
$\hat{Sh}_{i,j}  = 0$, $\widehat{Sh^2_{i,j}}  = 0$ and $N_{i,j} = 0 \text{ for } \{i, j\} \in \mathcal{\mathcal{I}}$.
\\
Stage 1. Pilot simulation to estimate the sample standard deviation:\\
\For{$n = 1,2,\ldots, N_0$}{
Sample model parameters $\pmb{w}^{(n)}$ and permutation $\pi^{(n)}$ from $p(\pmb{w}\vert\mathcal{D})$ and $\Pi(\mathcal{I})$;\\ 
\For{all set $\{i, j\}\subseteq \mathcal{I}$}{
$N_{i,j} = N_{i,j} + 1$;\\
$\Delta^{(n)}_{i,j}=\hat{g}_{nn}(P_{i,j}(\pi^{(n)})\cup \{i, j\}\vert\pmb{w}^{(n)})  - \hat{g}_{nn}(P_{i,j}(\pi^{(n)})\cup\{i\}\vert\pmb{w}^{(n)})- \hat{g}_{nn}(P_{i,j}(\pi^{(n)})\cup\{j\}\vert\pmb{w}^{(n)})+\hat{g}_{nn}(P_{i,j}(\pi^{(n)})\vert\pmb{w}^{(n)})$;\\
$\hat{\mbox{Sh}}_{i,j}=\hat{\mbox{Sh}}_{i,j}+\Delta^{(n)}_{i,j}$, $\widehat{Sh^2_{i,j}}=\widehat{Sh^2_{i,j}}+(\Delta^{(n)}_{i,j})^2$;
}}
$\hat{\sigma}_{i,j} = \sqrt{\dfrac{\widehat{Sh^2_{i,j}}+(\hat{Sh}_{i,j})^2/N_0}{N_0-1}}$;\\
Stage 2. Sequential allocation of the sample budget:\\
\For{$n = N_0+1,N_0+2,\ldots,N$}{
Calculate the gradient with respect to the half length $\mbox{gHL}_{i,j}$ 
for all $\{i,j\}\subseteq \mathcal{I}$;\\
Sample model parameters $\pmb{w}^{(n)}$ and permutation $\pi^{(n)}$ from $p(\pmb{w}\vert\mathcal{D})$ and $\Pi(\mathcal{I})$;\\
Find $\{i^*,j^*\} = \mbox{argmax}_{i,j}\mbox{gHL}_{i,j}$ and $i^*$ that is the leftmost element of $\{i^*,j^*\}$ in $\pi^{(n)}$;\\ 
Construct the group $G_m(i^*, \pi^{(n)})$;\\
\For{all set $\{i, j\}\subseteq G_m(i^*, \pi^{(n)})$}{
$N_{i,j} = N_{i,j} + 1$;\\
$\hat{\mbox{Sh}}_{i,j}=\hat{\mbox{Sh}}_{i,j}+\hat{g}_{nn}(P_{i,j}(\pi^{(n)})\cup \{i, j\}\vert\pmb{w}^{(n)})  - \hat{g}_{nn}(P_{i,j}(\pi^{(n)})\cup\{i\}\vert\pmb{w}^{(n)})- \hat{g}_{nn}(P_{i,j}(\pi^{(n)})\cup\{j\}\vert\pmb{w}^{(n)})+\hat{g}_{nn}(P_{i,j}(\pi^{(n)})\vert\pmb{w}^{(n)})$.
}}
Stage 3. Calculate the Shapley interaction effect: $\hat{Sh}_{i,j} = \hat{Sh}_{i,j}/N_{i,j} \text{ for } \{i,j\} \in \mathcal{I}$.
\caption{Shapley Interaction Effect Estimation with Sequential Budget Allocation}\label{agm: svsequential}
\end{algorithm}	

In Proposition~\ref{thm: CI for sv}, the half length of the asymptotic confidence interval $\mbox{HL}_{i,j} = z_{\alpha/2}\hat{\sigma}_{i,j}/\sqrt{N_{i,j}}$ quantifies the estimation uncertainty due to finite sample size. The corresponding absolute value for the gradient of $\mbox{HL}_{i,j}$ with respect to the number of samples $N_{i,j}$ is 
$\mbox{gHL}_{i,j} = \dfrac{z_{\alpha/2}\hat{\sigma}_{i,j}}{2N_{i,j}^{3/2}}$,
which reflects the benefit of one additional
sample in estimating $\hat{Sh}_{i,j}$. Therefore, instead of allocating the simulation budget to each $\{i,j\}$ uniformly, one can sequentially allocate the simulation budget to $\{i^*,j^*\}$ with the greatest
$\mbox{gHL}_{i,j}$ at each iteration until the simulation budget is exhausted.

As we described in Section~\ref{sec: shapeffect}, Algorithm~\ref{agm: svgeneral}
can benefit from the reused calculation of the value function. To take advantage of the reused calculation, we can modify the sequential procedure as follows: Without loss of generality, let 
$G(i^*, \pi^{(n)}) \subseteq \mathcal{I}$ denote the group of subsets with cardinality of 2 whose first element is $i^*$. At each iteration, we allocate the simulation efforts to $\{i^*,j\}$s that belong to $G(i^*, \pi^{(n)})$ and have top $m$ \mbox{gHL} among $G(i^*, \pi^{(n)})$, which is denoted by $G_m(i^*, \pi^{(n)})$. These $\{i^*,j\}$s have a common item $\hat{g}(P_{i^*,j}(\pi^{(n)})\vert \pmb{w}^{(n)})-\hat{g}(P_{i^*,j}(\pi^{(n)})\cup\{i^*\}\vert \pmb{w}^{(n)})$ when estimating the value function, which can be calculated once and then recorded to reduce the computation load. Algorithm~\ref{agm: svsequential} presents the pseudo-code of the sequential budget allocation procedure for Shapley interaction effect estimation. 

To further reduce the estimation error
of the interaction effect estimation, we apply the quasi-Monte Carlo method in Algorithm~\ref{agm: svsequential}, which achieves a lower estimation error
of the Shapley-Owen index compared to the standard Monte Carlo method by careful selection of samples.
We first generate low discrepancy point sets (e.g., Halton sequence), then map $n$th low discrepancy point $\pmb{x}^{(n)}$ to an actual sample $\{\pmb{w}^{(n)}$, $\pi^{(n)}\}$. More specifically, let $d_w$ denote the dimension of $\pmb{w}^{(n)}$ and $d_\pi$ denote the dimension of $\pi^{(n)}$, the size and dimension of the generated low discrepancy point sets be $N$ and $d_w+d_\pi -2$, respectively. The first $d_w$ dimensions of $x^{(n)}$ are mapped to $w^{(n)}$ according to \shortciteN{Lemieux2009} 
and the remaining $(d_\pi -2)$ dimensions are mapped to $\pi^{(n)}$ using the transformation method in \shortciteN{mitchell2022sampling}. 

\section{Numerical Experiment}
\label{sec: experiment}

In this section, we present the numerical results for two types of PABN where we estimate the Shapley-Owen indices. The first one is the linear Gaussian PABN in \shortciteN{zheng2023policy}, which is valid for bioprocesses with online monitoring, meaning that the monitoring frequency is on a faster time scale than the evolution of bioprocess dynamics. As the value function $g(\cdot)$ for linear PABN can be analytically calculated, we can obtain the Shapley-Owen indices precisely. Therefore, we use linear PABN to
test the performance of the Algorithm~\ref{agm: svgeneral} and~\ref{agm: svsequential} on estimating
the Shapley-Owen indices. In addition, we conduct the ablation study on the choice of sample size at each level of Algorithm~\ref{agm: svgeneral} and the impact of the sampling methods (Monte Carlo and quasi-Monte Carlo) in Algorithm~\ref{agm: svsequential}. The second one is a PABN application with nonlinear transition functions that describe the general mature cell-progenitor negative feedback process. We analyze the estimated value of the Shapley-Owen indices for this real-world PABN and demonstrate the effectiveness of these indices in capturing interaction effects and providing guidance for decision-making. 

\subsection{Linear Gaussian PABN Model}
For the linear Gaussian PABN model, we consider the case that the transition function has a linear form and the random factors follow a Gaussian distribution.  More accurately, let $\pmb\beta_{t}^s$ denote the $d_s\times d_s$ matrix whose $\left(n, q\right)$-th element is the linear coefficient $\beta^{nq}_t$ 
corresponding to the effect of state $s^n_t$ on the next state $s^q_{t+1}$. 
Similarly, let $\pmb\beta_t^a$ be the $d_a\times d_s$ matrix representing the effects of each component of $\pmb{a}_t$ on each component of $\pmb{s}_{t+1}$. Then the dynamic function becomes
    $\pmb{s}_{t+1} = \pmb{\mu}_{t+1}^s + \left(\pmb{\beta}_{t}^s\right)^\top(\pmb{s}_t-\pmb\mu_t^s) + \left(\pmb\beta_{t}^a\right)^\top(\pmb{a}_t-\pmb\mu_t^a) +\pmb{e}_{t+1}$, 
where $\pmb\mu_{t}^s=(\mu_t^{1},\ldots, \mu_t^{d_s})$, $\pmb\mu_{t}^a=(\lambda_t^{1},\ldots, \lambda_t^{d_a})$ and $\pmb{e}=(\pmb{e}_1^\top, \pmb{e}_2^\top,\ldots,\pmb{e}_H^\top)\sim \mathcal{N}(\pmb{0}, \mathbf{V})$.
The
model parameters $\pmb{w} = (\pmb{\mu}^s,\pmb\mu^a,\pmb{\beta}, \pmb{V})$, where $\pmb{\mu}^s = \{\pmb{\mu}_{t}^s\}_{t = 1}^{H}$, $\pmb{\mu}^a = \{\pmb{\mu}_{t}^a\}_{t = 1}^{H-1}$, $\pmb\beta = \{(\pmb\beta_t^a,\pmb\beta_t^s)\}_{t=1}^{H-1}$. 
The policy and reward functions are $\pmb{a}_t = \pmb\mu^a_t + \pmb{\theta}_t^\top(\pmb{s}_t - \pmb\mu^{s}_t)$ and $r_t\left(\pmb{s}_t,\pmb{a}_t\right) = m_t + \pmb{b}^\top_t\pmb{a}_t + \pmb{c}^\top_t\pmb{s}_t$,
where $\pmb{\theta}_t$ is an $d_s\times d_a$ matrix. 

Given model parameters $\pmb{w}$, the cumulative reward $r(\pmb\theta\vert\pmb{w})$ can be written as
$r\left(\pmb\theta\vert\pmb{w}\right)=  \sum^{H}_{t=1}r_t(\pmb{s}_t, \pmb{a}_t\vert\pmb w)
=\gamma
+\sum^{H}_{t=1}\pmb{\alpha}_t\left(\sum_{t'=1}^{t}\mathbf{R}_{t',t-1}\pmb{e}_{t'}\right)$,
where $\gamma$ denotes the term that excludes random factors,
$\pmb{\alpha}_t =\pmb{b}_t^\top\pmb\theta_t^\top+\pmb{c}_t^\top$, $\mathbf{R}_{t',t} =\prod_{t^{''}=t'}^t\left[\left(\pmb\beta_{t^{''}}^s\right)^\top + \left(\pmb\beta_{t^{''}}^a\right)^\top\pmb\theta_{t^{''}}^\top\right]$ represents the product of pathway coefficients from time step $t'$ to $t$ and $\mathbf{R}_{t',t'-1} = \mathbb{I}_{d_s\times d_s}$, the $d_s\times d_s$ identity matrix. Let $\pmb{R}_{t'} = \sum^{H}_{t=t'}\pmb{\alpha}_t\mathbf{R}_{t',t-1}$, the variance of $r\left(\pmb\theta\vert\pmb{w}\right)$ is $
\mbox{Var}\left[r(\pmb{\theta}\vert\pmb{w})\right]= \mbox{Var} \left[\sum^{H}_{t=1} \pmb{\alpha}_t\left(\sum_{t'=1}^{t}\pmb{R}_{t',t-1}\pmb{e}_{t'}\right)\right]=\sum^{H}_{t_1=1}\sum^{H}_{t_2=1}\left[\pmb{R}_{t_1}\mbox{Cov}(\pmb{e}_{t_1},\pmb{e}_{t_2})\pmb{R}_{t_2}^{\top}\right]$.

\begin{table}[!b]
\centering
\caption{MSE of Algorithm~\ref{agm: svgeneral} using different budget allocation ratios with 90\% confidence interval.}
\label{tab: NI performance}
\begin{tabular}{cllllll}
\toprule
\multirow{2}{*}{$K$:$M$:$N_O$} & \multicolumn{5}{c}{Number assigned to $N_I$ (increasing order)}  \\
\cmidrule(lr){2-6}
 & 1 & 2 & 3 & 4& 5\\ \midrule
1:1:1& 20.9 (3.9)& 26.9 (8.6) & 28.3 (7.1) & 67.2 (17.0)& \textbf{81.4 (18.5)} \\
1:3:5& 17.2 (2.4)& 19.2 (3.6) & 27.9 (5.5) & 28.8 (7.3)& \textbf{40.6 (7.8)}\\
2:3:6 & 17.5 (5.1) & 19.1 (4.2) & 22.5 (6.1) & 28.9 (4.9) & \textbf{44.9 (8.3)}\\
6:2:4 & 18.6 (1.9) & 20.4 (5.1) & 23.9 (4.3) & 28.6 (4.0) & \textbf{68.2 (13.3)}\\
4:3:3 & 16.9 (2.9) & 18.3 (3.4) & 22.9 (4.2) & 25.8 (8.5) & \textbf{49.1 (7.2)}\\
1:6:1 & 16.9 (1.3) & 19.0 (5.1) & 22.1 (7.2) & 40.3 (4.0)& \textbf{44.3 (10.2)}\\
5:4:3& 20.7 (3.2) & 25.0 (7.2) & 26.2 (4.5) & 36.6 (8.1)& \textbf{48.3 (9.5)} \\
2:1:6& 18.3 (2.1) & 26.0 (8.0) & 27.8 (7.8) & 69.9 (14.0) & \textbf{90.0 (7.8)}\\
5:5:1& 17.2 (3.7) & 21.4 (3.2) & 26.0 (6.3) & 39.7 (7.9) & \textbf{94.7 (22.2)} \\
3:1:3& 18.8 (1.7) & 24.3 (5.2) & 40.5 (10.9) & 45.2 (7.4)&  \textbf{62.4 (25.7)}\\
\bottomrule
\end{tabular}
\end{table}

Now we calculate the value function for the linear Gaussian PABN. Let $\pmb{R} = (\pmb{R}_{1}, \pmb{R}_{2},\ldots,\pmb{R}_{H})$, the cumulative reward can be written as 
$r\left(\pmb\theta\vert\pmb{w}\right)
=\gamma+\pmb{R}\pmb{e}$. 
If we denote
$\pmb{R}_{\mathcal{U}}$ as the corresponding coefficient vector in  $\pmb{R}$ for input set $\mathcal{U}$
, and similarly, let $\Sigma_{-\mathcal{U}, \mathcal{U}}$ be the the covariance matrix between random inputs not in $\mathcal{U}$ and random inputs in $\mathcal{U}$,  $\Sigma_{\mathcal{U}}$ denote the covariance matrix between random inputs in $\mathcal{U}$, then we can derive the value function $g(\mathcal{U}\vert\pmb{w})$ explicitly as follows:
\begin{equation}
\label{eq: linear value function}
    \begin{aligned}
        g(\mathcal{U}\vert\pmb{w}) 
        =(\pmb{R}_{-\mathcal{U}}\Sigma_{-\mathcal{U},\mathcal{U}}\Sigma_{\mathcal{U}}^{-1}+\pmb{R}_{\mathcal{U}})\Sigma_{\mathcal{U}}(\pmb{R}_{-\mathcal{U}}\Sigma_{-\mathcal{U},\mathcal{U}}\Sigma_{\mathcal{U}}^{-1}+\pmb{R}_{\mathcal{U}})^{\top}.
    \end{aligned}
\end{equation}

\begin{figure}[!t]
\vspace{-1em}
    \centering    \includegraphics[width=0.6\textwidth]{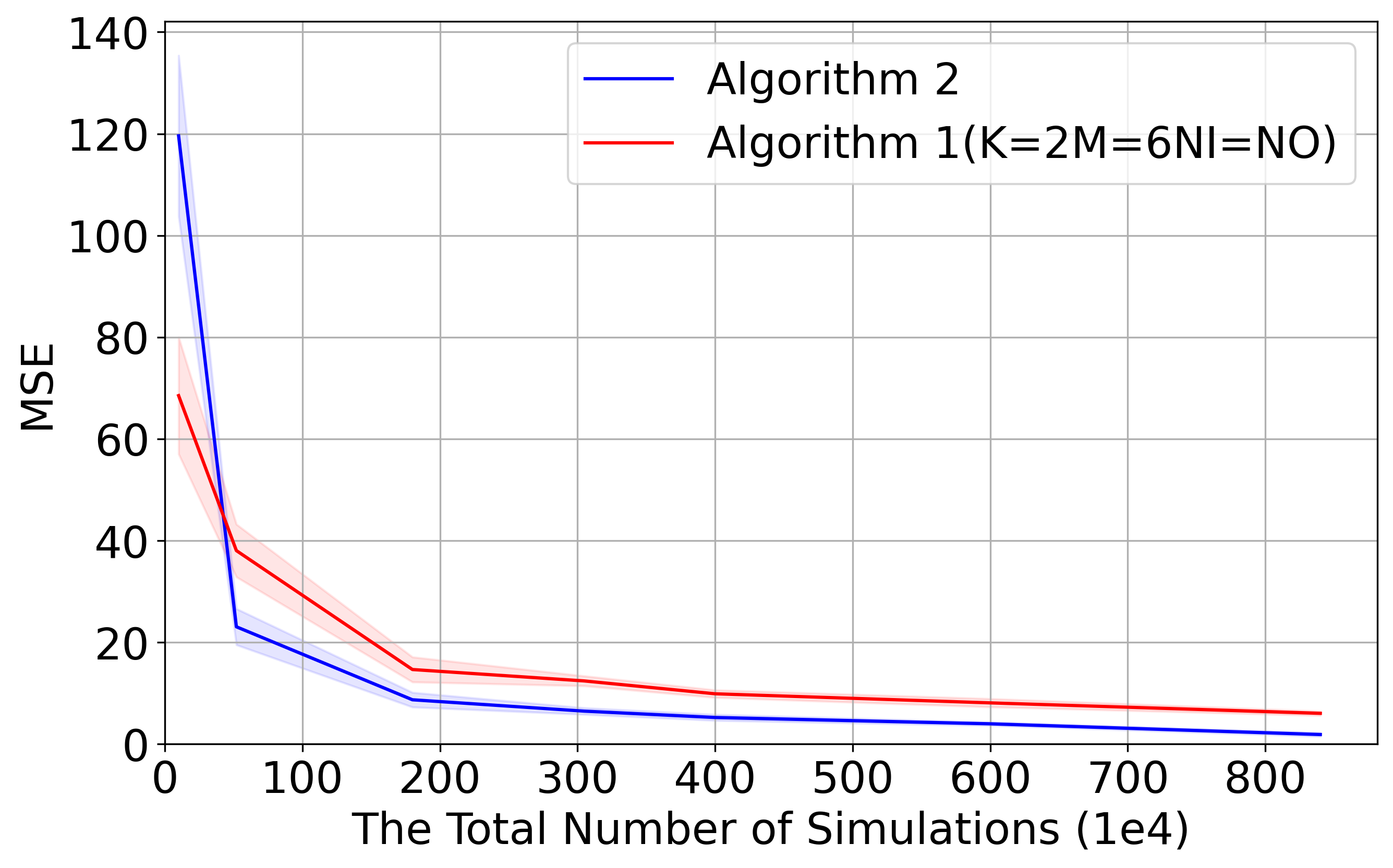} 
    \caption{MSE of Algorithm~\ref{agm: svgeneral} and Algorithm~\ref{agm: svsequential}}
    \label{fig: agm1 and agm2}
\end{figure}

In our experiments, we set the number of states $d_s = 3$, the number of actions $d_a = 1$, and the time horizon $T = 2$. This configuration results in 6 factors and 15 interactions. The posterior distribution of the model parameters $p(\pmb{w}\vert\mathcal{D})$ is set to be multivariate Gaussian. We first test the performance of Algorithm~\ref{agm: svgeneral} with different budget allocation strategies. Following \shortciteN{hart2019global}
, we calculate the mean squared error (MSE) to measure the performance of the algorithm, which is defined as
$        \mbox{MSE}=\frac{1}{15}\sum_{\{i,j\}\subseteq \mathcal{I}}(\mbox{Sh}_{i,j}-\hat{\mbox{Sh}}_{i,j})^2$,
where $\mbox{Sh}_{i,j}$ is the real Shapley-Owen value estimated by using a large number of model parameter samples and calculating the value function with Equation~(\ref{eq: linear value function}), $\hat{\mbox{Sh}}_{i,j}$ is obtained from Algorithm~\ref{agm: svgeneral}.


We first conduct ablation study to determine the sample size for each level of simulation given a fixed computational budget, i.e., to decide $K$, $M$, $N_O$ and $N_I$. As we cannot find any literature that discusses the optimal allocation strategy for such four-level nested simulation, we apply the grid search and assign a number in ${1,2,3,4,5,6}$ to each level of simulation. Fixing $K*M*N_O*N_I = 30000$, we use the ratio of 4 levels (i.e., $K:M:N_O:N_I$) to allocate the simulation budget. We find that the budget allocation ratio with the smallest MSE is $K:M:N_O:N_I = 6:3:6:1$, which means a good budget allocation policy tends to bound the number of inner level samples $N_I$. In order to validate this conclusion, we randomly sample 10 different budget allocation ratios for $K$, $M$ and $N_O$ (i.e., $K:M:N_O$), and for each ratio, we vary the number assigned to $N_I$ and record the corresponding MSE averaged by 10 macro-replications in Table~\ref{tab: NI performance}. We find that the MSE of Algorithm~\ref{agm: svgeneral} increases as the proportion of inner sample size $N_I$ increases, which is consistent with the conclusion in \shortciteN{sun2011efficient} that, as the computational budget increases, the optimal number of inner-level samples remains bounded.

\begin{table}[!t]
\centering
\caption{MSE of algorithm 2 using MC and QMC implementation with 90\% confidence interval.}
\label{tab: mcqmc}
\begin{tabular}{cccccc}
\toprule
Sample size& 180000 & 360000 & 540000 & 720000 & 900000  \\ \midrule
MC & 110.2 (9.0) &  62.4 (4.7) & 26.0 (2.4) & 22.7 (1.9) & 20.5 (1.8) \\
QMC & 87.1 (7.7) & 49.9 (3.9) & 20.5 (2.1) & 17.7 (1.9) & 16.2 (1.3)\\

\bottomrule
\end{tabular}
\end{table}

Next, we compare the performance of Algorithm~\ref{agm: svgeneral} and Algorithm~\ref{agm: svsequential}. For Algorithm~\ref{agm: svgeneral}, we use the budget allocation policy $K:M:N_O:N_I = 6:3:6:1$ and for Algorithm~\ref{agm: svsequential}, we set $N_0 = 0.02N$ and $m = 2$. The result is illustrated in Figure~\ref{fig: agm1 and agm2}, which indicates that the performance of Algorithm~\ref{agm: svsequential} is superior over Algorithm~\ref{agm: svgeneral} except when the budget is very limited. This is because the estimation of the variance might be inaccurate without a sufficient simulation budget. Finally, we compare the performance of Algorithm~\ref{agm: svsequential} using classical Monte Carlo (MC) and quasi Monte Carlo (QMC) sampling methods. As shown in Table~\ref{tab: mcqmc}, we can see the QMC can significantly reduce the estimation error of Algorithm~\ref{agm: svsequential}.


\subsection{Nonlinear PABN Model}

For the nonlinear PABN model, we use a general mature cell-progenitor negative feedback model in \shortciteN{stacey2018experimentally} as an example to interpret the results of Shapley-Owen indices. In this system, cytokine-mediated negative feedback suppresses the bulking up of progenitors in the presence of mature cells, and the productivity of a volume of medium is therefore determined by the accumulation of inhibitory cytokines. The transition model that characterizes the evolution of progenitor density (S), product density (P) and inhibitor concentration (I) over time can be described as 
$\frac{dS}{dt}=\frac{r_g}{(1+\exp(a(I-b)))}-r_cS$, 
     $\frac{dP}{dt} = r_cS-r_dP$,   $\frac{dI}{dt} =  r_pP$,
where $r_g$ is growth rate, $r_c$ is conversion rate from progenitor to product, $r_d$ is product death rate, $r_p$ is production rate of inhibitor, $a$ is inhibitor sensitivity and $b$ is inhibitor threshold. Therefore, the state vector in this system is $\pmb{s}_t = \{S_t, P_t, I_t\}$.
The values of these parameters are set according to \shortciteN{stacey2018experimentally}. By reducing the concentration of inhibitor by a media dilution, there is the potential to increase productivity.  Therefore, the action in this system is the fraction of the concentration reduction of $I$ at each period. For illustration, we simply adopt the policy that we dilute the concentration to be half of the current concentration at each decision time. The output of interest is the reward of final product density minus the cost of the inhibitor concentration reduction.

Let the random factors associated with $S$, $P$ and $I$ at period $t$ be $e^S_t$, $e^P_t$, and $e^I_t$. Following  \shortciteN{mcadams1997variations}, the culture PH can be negatively related to the progenitor density and positively related to the product density and inhibitor concentration over a specific range, we set $e^i_t = l^{i} e^{PH}_t + e^{i'}_t$ for $i\in \{S, P, I\}$, where $e^{PH}_t$ denotes the randomness associated with the culture PH, $l^{i}$ is a coefficient whose sign indicates the correlation relationship, and $e^{i'}_t$ denotes other sources of randomness. 
We vary the values of $\mathbf{l} = \{l^{S}, l^{P}, l^{I}\}$ as "no dependence" ($\mathbf{l} = \{0, 0, 0\}$), "low dependence" ($\mathbf{l} = \{-0.5, 0.5, 0.5\}$), "strong dependence" ($\mathbf{l} = \{-1, 1, 1\}$) to show the Shapley-Owen indices at different level of correlation. 

\begin{figure}[!t]
\vspace{-1em}
	\centering	\includegraphics[width=0.7\textwidth]{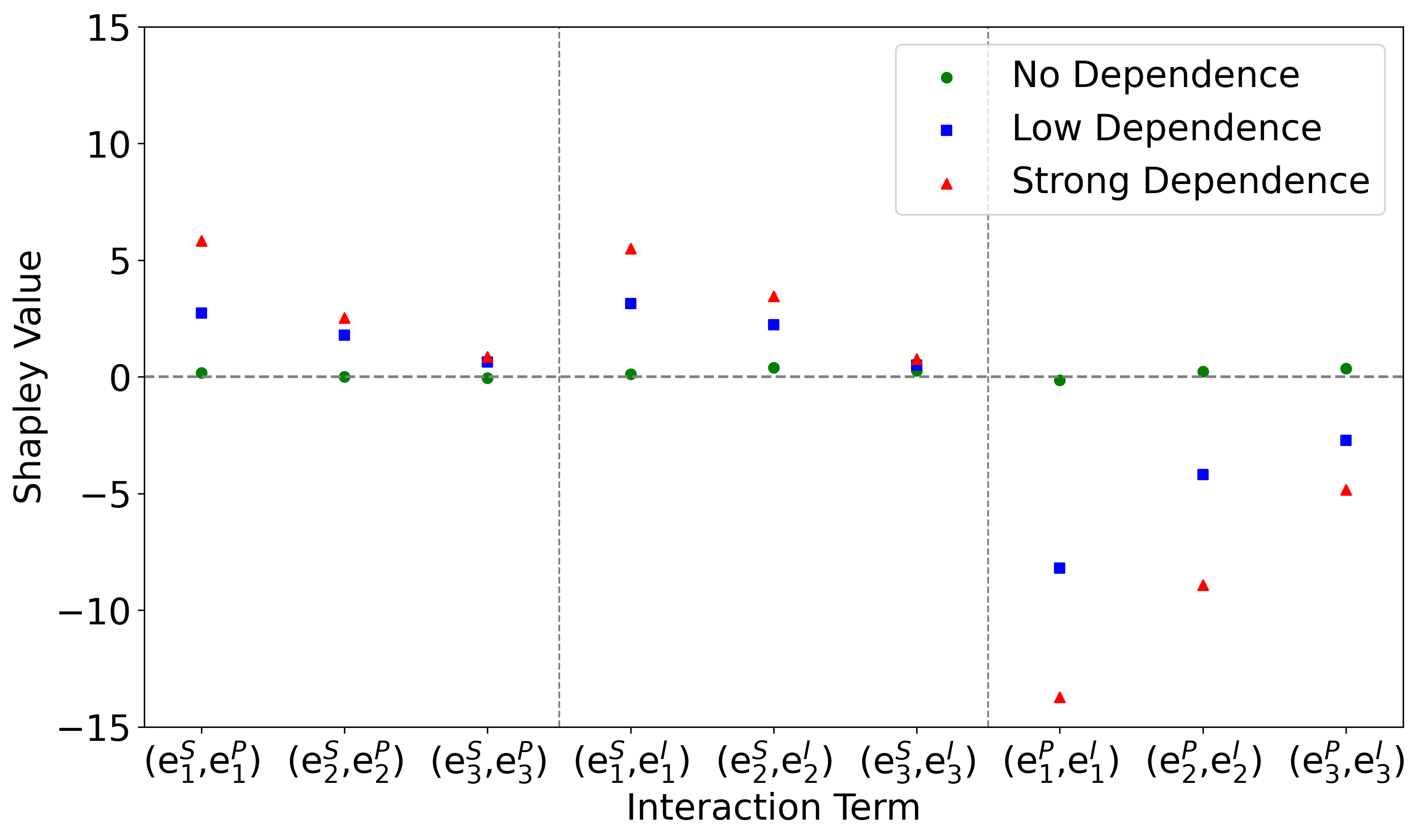}
\caption{Shapley interaction effects for mature cell-progenitor negative feedback model.}
	\label{fig: nonlinear}
\end{figure}
The estimated Shapley-Owen index is shown in Figure~\ref{fig: nonlinear}. First, it is evident that the interaction effects of ($S$, $P$), ($S$, $I$) and ($P$, $I$) diminish as time proceeds. This is because the interactions at the early stage can be amplified and accumulated to impact the final output through the nonlinear transition model. Second, we find that as the level of correlations grows, the magnitude of the Shapley-Owen indices also increases to capture the interaction effects on the output variance, which shows the power of Shapley-Owen indices in measuring the interaction effects. Notice that such interactions among the random factors often occur due to the underlying factors. Therefore, a systematic experiment on factors with large interaction effects can help detect the underlying factors that contribute greatly to the system variation. Third, under the scenario of strong dependence, we find that the introduced correlation between the random factors associated with $P$ and $I$ can reduce the output variance. Therefore, a sophisticated control of the product density based on the inhibitor concentration can help improve the process stability. For example, \shortciteN{liu2015butanol} suggested that one can use activated carbon to control the concentration of inhibitors during butanol production and show that it can yield high butanol production and reduce the output variation. 

\section{Conclusion}
\label{sec: conclusion}
In this paper, we present a method for conducting a global sensitivity analysis on the interaction effects of the PABN model in biomanufacturing. We use the Shapley-Owen index to quantify the interaction effects and introduce a sampling-based framework to approximate the interaction measures. To enhance computational efficiency, we further incorporate a non-nested simulation method and a sequential budget allocation scheme. Numerical results show that our algorithms have superior empirical performance in terms of accuracy and efficiency. In future research, it will be worth considering combining the variance reduction techniques tailored to the quantification of interaction effects for PABN models that involve high noise. In addition, how to use both the main effects and the interaction effects for factor screening in biomanufacturing is also an interesting problem.

\section*{ACKNOWLEDGMENTS}
The first, second, and fourth authors were supported in part by the National Natural Science Foundation of China (Grants 72031006, 72394370, 72394375).

\bibliographystyle{wsc}

\bibliography{wscref}

\section*{AUTHOR BIOGRAPHIES}

\noindent {\bf JUNKAI ZHAO}
 is a Ph.D. student in the Antai College of Economics and Management 
at Shanghai Jiao Tong University. His research interest is simulation optimization. His email address is \email{zhaojunkai@sjtu.edu.cn}.\\


\noindent {\bf JUN LUO}
is a professor in the Antai College of Economics and Management at Shanghai Jiao Tong
University. His primary research interests are stochastic simulation and simulation optimization. His email address is \email{jluo\_ms@sjtu.edu.cn}.\\

\noindent {\bf WEI XIE}
is an assistant professor in Mechanical and Industrial Engineering at Northeastern University. Her research interests include AI/ML, computer simulation, data analytics, and stochastic optimization. Her email address is \email{w.xie@northeastern.edu}. 
\\

\noindent {\bf ZIXUAN BAI}
is a master student in the Antai College of Economics and Management at Shanghai Jiao Tong University. His email address is \email{zixuanbai@sjtu.edu.cn}.\\
\end{document}